\documentclass[aps,prl,twocolumn,groupedaddress]{revtex4}
\usepackage{color}
\usepackage[dvips]{graphicx}
\def\beq{\begin{equation}}
\def\eeq{\end{equation}}
\newcommand{\capa}{\mathrm{A}}
\newcommand{\capb}{\mathrm{B}}

\begin{document}


\title{Successive phase transitions at finite temperatures of the supersolid in the three-dimensional extended Bose-Hubbard model}


\author{Keisuke Yamamoto,${}^1$ Synge Todo,${}^{2,3}$ and Seiji Miyashita${}^{1,3}$}
\affiliation{${}^1$Department of Physics, School of Science, The University of Tokyo, 7-3-1 Hongo, Bunkyo-ku, Tokyo 113-0033, Japan,\\
${}^2$Department of Applied Physics, The University of Tokyo, 7-3-1 Hongo, Bunkyo-ku, Tokyo 113-8656, Japan,\\
${}^3$CREST, JST, 4-1-8 Honcho Kawaguchi, Saitama, 332-0012, Japan. }

\collaboration{}

\date{\today}

\begin{abstract}
We study the finite temperature properties of the extended Bose-Hubbard model on a cubic lattice. 
This model exhibits the so-called supersolid state.
To start with, we investigate ordering processes by quantum Monte Carlo simulations, and find successive superfluid and solid phase transitions. 
There, we find that the two order parameters compete with each other.
We obtain the finite temperature phase diagram, which contains the superfluid, the solid, the supersolid and the disordered phase.
We develop a mean-field theory to analyze the ordering processes and compare the result with that obtained by simulations,
and discuss the mechanism of the competition of these two orders.
We also study how the supersolid region shrinks as the on-site repulsion becomes strong.  

\end{abstract}

\pacs{}

\maketitle

\section{Introduction}
The supersolid state is an interesting state of matter which has both solid and superfluid properties.
The solid state is characterized by the breaking of translational symmetry, and the superfluid state is characterized by the breaking of $U(1)$ symmetry of the phase of macroscopic wave function.
Thus, the simultaneous breaking of these two symmetries indicates that there is a flow component in solid. 
The possibility of the supersolid was first discussed by Penrose and Onsager \cite{penrose1956}.
Since then, various studies on supersolid have been conducted from both experimental and theoretical points of view. 

As to observation of supersolid, Leggett suggested that non-classical rotational inertia (NCRI) would be available to detect supersolid in rotating solid ${}^4$He.
Recently, Kim and Chan \cite{kim2004}, reported that they found NCRI in solid $^4$He.
Although it has been pointed out that the observed NCRI may not due to the supersolid but due to the grain boundaries between polycrystals \cite{sasaki2006},
the topic, however, still attracts researchers' interest.

The possibility of supersolid on lattice models has been discussed actively.
Andreev and Lifshitz \cite{andreev1969}, suggested that the delocalization of the vacancies in crystal causes a mass flow.
Matsuda and Tsuneto \cite{matsuda1970}, studied the ground state of the hardcore Bose-Hubbard model using mean-field theory, and they showed that the supersolid 
is possible when the interaction of particles has frustration.
This is confirmed by numerical simulations in two-dimensional \cite{boninsegni2003,batrouni2000}, and three dimensional cases \cite{suzuki2007}.

It has been pointed out that the supersolid state can be realized even in non-frustrated lattice if double occupancy of the particles is allowed \cite{sengupta2005}.  
In the case of a square lattice, the supersolid is found in the ground state when the number density of particle is more than that of the half-filled case $(\rho>1/2)$
 \cite{sengupta2005}.
Although ground state properties of the supersolid have been studied extensively, there is no direct study on the finite temperature properties of supersolid problem
in the extended Hubbard model.

In this paper, we study finite temperature properties of the supersolid state in the extended Bose-Hubbard model on a cubic lattice.
In our model, the hopping term causes the superfluid order and the nearest neighbor repulsive interaction tends to form the solid order.
These orders compete with each other, and in the hardcore limit, these orders cannot be realized simultaneously. Thus, there is no supersolid state.  
However, as has been pointed out \cite{sengupta2005}, these orders can be realized together in the soft-core case even in not-frustrated lattices such as a square lattice.
We investigate ordering processes of superfluid and solid at finite temperatures in a cubic lattice, using the stochastic series expansion (SSE) method.
As a result, we have a phase diagram of superfluid, normal solid, disordered state, and supersolid. 
We also study the finite temperature dependence of the orders by making use of mean-field (MF) analysis
 and compare the result with that obtained by SSE. They show qualitatively good agreement.
Moreover, the competition of the solid and superfluid order is discussed using Ginzburg-Landau free energy.
Finally, we study the effect of on-site repulsion $U$ on the coexistence of the two orders. 
We find how the supersolid region at finite temperature shrinks as $U$ becomes large. 

\section{Model}
We analyze the extended Bose-Hubbard Hamiltonian on a cubic lattice
\begin{eqnarray}
 {\cal H}=-t\sum_{<ij>}(a_i^{\dagger}a_j +a_i a_j^{\dagger}) +V\sum_{<ij>}n_i n_j \nonumber \\
+\frac{1}{2} U \sum_{i} n_i (n_i-1)- \mu \sum_{i} n_i ,
 \label{eq-hamiltonian}
\end{eqnarray}
where $ a_i^{\dagger}$ and $a_i $ are the creation and annihilation operators of a boson $([ a_i, a_j^{\dagger}]=\delta_{ij})$, and
$n_i= a_i^{\dagger}a_i $. The parameter
$t$ denotes the hopping matrix element, $U$ and $V$ are the on-site and nearest neighbor repulsion, respectively, and $\mu$ is the chemical 
potential. The notation $\langle ij \rangle$ means the sum over the nearest neighbor pairs.
The system size  is $N=L^3$, where $L$ is the length of the system.
The order parameter of the solid state is
\begin{eqnarray}
S_{\pi}=\frac{1}{N^2}\sum_{jk} e^{i \mathbf{Q} \cdot (\mathbf{r}_j-\mathbf{r}_k)}\langle n_j n_k \rangle, \label{eq-defSpi}
\end{eqnarray}
where $\mathbf{Q}=(\pi,\pi,\pi)$ is the wave vector that represents the staggered order.
The order parameter of the superfluid state is
\begin{eqnarray}
\rho_{s}=\frac{1}{N^2}\sum_{jk} \langle a_j^{\dagger} a_k +a_k^{\dagger} a_j \rangle , \label{eq-defRhos}
\end{eqnarray}
which represents the off diagonal long range order (ODLRO).

\section{Methods}
We use the following two different methods to analyze properties of the system.
\subsection{(1) Stochastic Series Expansion}
We preform numerical simulation of the stochastic series expansion (SSE),
which was invented by Sandvik \cite{sandvik1991,sandvik1999}.
This method is one of quantum Monte Carlo (QMC) simulations, and has been successfully applied for various quantum systems. 
In order to avoid the clusterization due to diagonal frustration, we adopt the generalized directed loop algorithm \cite{alet2005}.  
We use a package of the ALPS \cite{alps2}.
We adopt a simple cubic lattice of $N=L^3$ sites with periodic boundary conditions along all the lattice axes.
In the simulation, we estimate the
superfluidity $\rho_s$, Eq. (\ref{eq-defRhos}), using the winding number $W$ of world lines defined by,
\begin{eqnarray}
\rho_s^{\mathrm{SSE}} = \frac{\langle W^2 \rangle}{3 t \beta L},\label{eq-defSSERhos}
\end{eqnarray}
which represents well $\rho_s$ \cite{pollock1987,prokofev2000}.
Here $\beta$ is the inverse temperature.  

\subsection{(2) Mean-Field Approximation}
We also analyze the ordering processes by the mean-field (MF) approximation \cite{xiancong2006}.
In order to study the solid state, we use a sublattice structure which is characterized by a staggered order of the density.
Here we adopt mean-fields for the solid order and superfluid order at sublattices A and B.
The Hamiltonian for this MF is given by

\begin{eqnarray}
{\cal H}_{\mathrm{MF}} &=& {\cal H}_{\capa}+{\cal H}_{\capb}+C \label{eq-meanham}, \\
{\cal H}_{\capa} &=& -zt(a_{\capa}^{\dagger}+a_{\capa}) \phi_{\capb} +zVn_{\capa} m_{\capb} \nonumber \\
                      &=& +\frac{U}{2} n_{\capa} (n_{\capa}-1) -\mu n_{\capa},  \\
{\cal H}_{\capb} &=& -zt(a_{\capb}^{\dagger}+a_{\capb}) \phi_{\capa} +zVn_{\capb} m_{\capa} \nonumber \\
                      &=& +\frac{U}{2} n_{\capb} (n_{\capb}-1) -\mu n_{\capb},   \\
C  &=& 2zt\phi_{\capa} \phi_{\capb} -zV m_{\capa} m_{\capb}, 
\end{eqnarray} 
where $z=6$ is the number of nearest neighbor sites.
Here, $m_{\capa}$ and $m_{\capb}$ are the mean fields correspond to the expectation values of the number operators for A and B sites, respectively Eq. (\ref{eq-self1}).
Similarly, $\phi_{\capa}$ and $\phi_{\capb}$ correspond to the expectation values of the annihilation operators for A and B sites, respectively.
\begin{eqnarray}
m_{\capa}=\langle n_{\capa} \rangle , &  m_{\capb}=\langle n_{\capb} \rangle , \label{eq-self1} \\
\phi_{\capa}=\langle a_{\capa} \rangle , &  \phi_{\capb}=\langle a_{\capb} \rangle . \label{eq-self2}
\end{eqnarray}
${\cal H}_{\capa}$ $({\cal H}_{\capb})$ is a mean-field Hamiltonian at a site of the A (B) sublattice.
$C$ is a correction term compensating the double counting of the energy.
The partition function and the free energy are given by
\begin{eqnarray}
{\cal Z}_{\mathrm{MF}} = \mathrm{Tr} (e^{-\beta {\cal H}_{\mathrm{MF}}}) \\
F_{\mathrm{MF}}= -\frac{1}{\beta} \mathrm{ln} {\cal Z}_{\mathrm{MF}}.
\end{eqnarray}
In MF, $m \equiv (m_{\capa}-m_{\capb})/2$ denotes the order parameter of solid 
and fulfill the relation, $S_{\pi}=m^2$.
Similarly, $ \phi \equiv (\phi_{\capa}+\phi_{\capb})/2 $ represents the order parameter of superfluid 
and fulfill the relation, $\rho_s=2\phi^2$.

\section{Ground State Properties} \label{sec-gs}

\begin{figure}
\begin{center}
\includegraphics[width=6cm,angle=270]{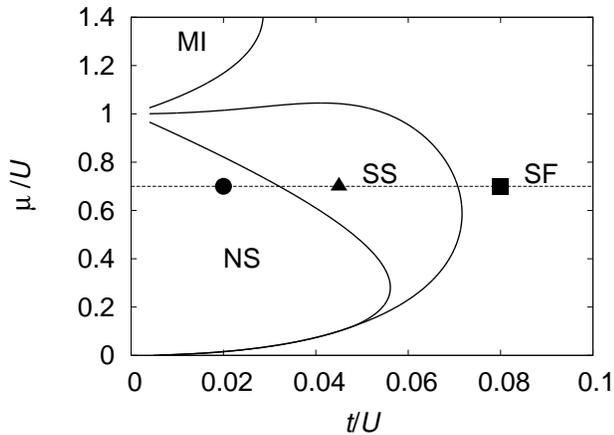}%
\caption{\label{fig-SST0}The ground state phase diagram of the soft-core Hubbard model for $V=U/z$ obtained by MF.
There are four phases: NS (normal solid), SS (supersolid), SF (superfluid), and MI (Mott-insulator).
Temperature dependence of order parameters calculated by SSE is given in Fig. \ref{fig-CBSFT} at the positions denoted by circle, triangle and rectangle. }
\end{center}
\end{figure}

Before analyzing properties at finite temperatures,
let us briefly summarize the ground state properties.
As has been reported, the system has the supersolid phase in the ground state when $U$ takes a finite value \cite{sengupta2005,otterlo1995}. 
In Fig. \ref{fig-SST0}, we show the ground state phase diagram in the coordinate of $(t/U, \mu/U)$ obtained by MF method.
Here we set $V=U/z$ at which the superfluidity takes maximum value \cite{YamamotoMiyashita1}.
The solid line is the phase boundary obtained by MF at $T/U=0.001$.
This ground state phase diagram agrees well with that obtained by the Gutzwiller variational method by Otterlo \cite{otterlo1995}.
In Fig.~\ref{fig-SST0}, we find four different phases, i.e. Mott-insulator (MI), normal solid (NS), superfluid (SF) and supersolid (SS). 
In addition to these four phases, a disordered phase representing the normal liquid (NL) phase appears at finite temperature.
We also study the ground states for several parameter sets by SSE. 
For example, we find the solid state for the set $(t/U=0.02, \mu /U=0.7)$ denoted by the circle in Fig. \ref{fig-SST0}, 
the superfluid state for $(t/U=0.08, \mu /U=0.7)$ by the rectangle, 
and the supersolid state for $(t/U=0.045, \mu /U=0.7)$ by the triangle.
The temperature dependences of order parameters on these points are given in the next section in Fig. \ref{fig-CBSFT} (a), (b) and (c).
We find the dependence on the parameters obtained by SSE agrees with that obtained by MF.

\section{Phase transitions at finite temperatures}
Now, we study the ordered states at finite temperatures.
The phase transition between the normal liquid phase and the solid phase belongs to the universality class of the Ising model.
The phase transition of superfluid belongs to the XY universality class.
In this section, we study the temperature dependence of these order parameters.
\subsection {Stochastic Series Expansion}
First, we show the results obtained by SSE. 
The simulations were performed in the grand canonical ensemble
using a system sizes $N=10^3$ and $N=12^3$.

\begin{figure}
\begin{center}
\includegraphics[width=5cm,angle=270]{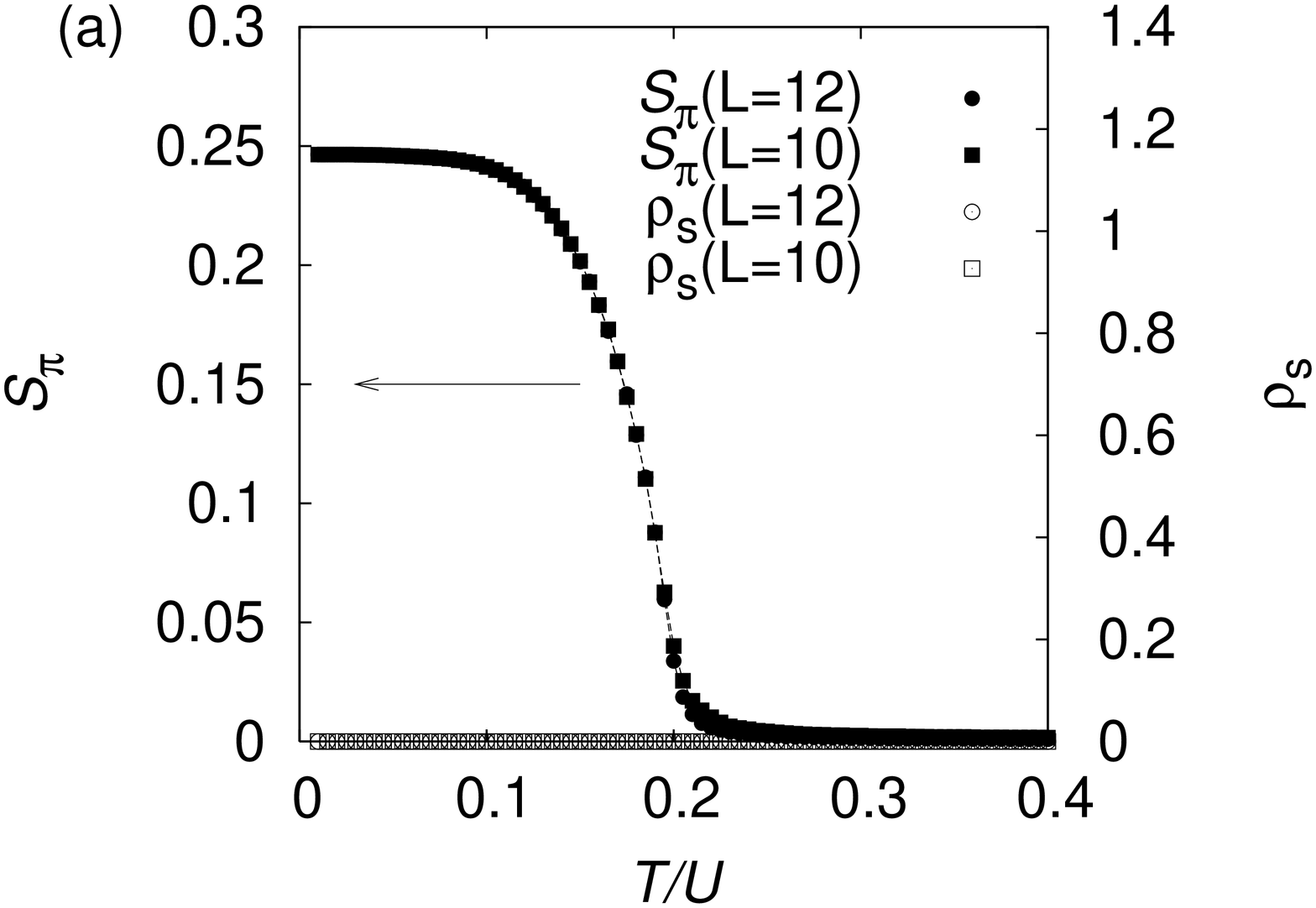} \\
\includegraphics[width=5cm,angle=270]{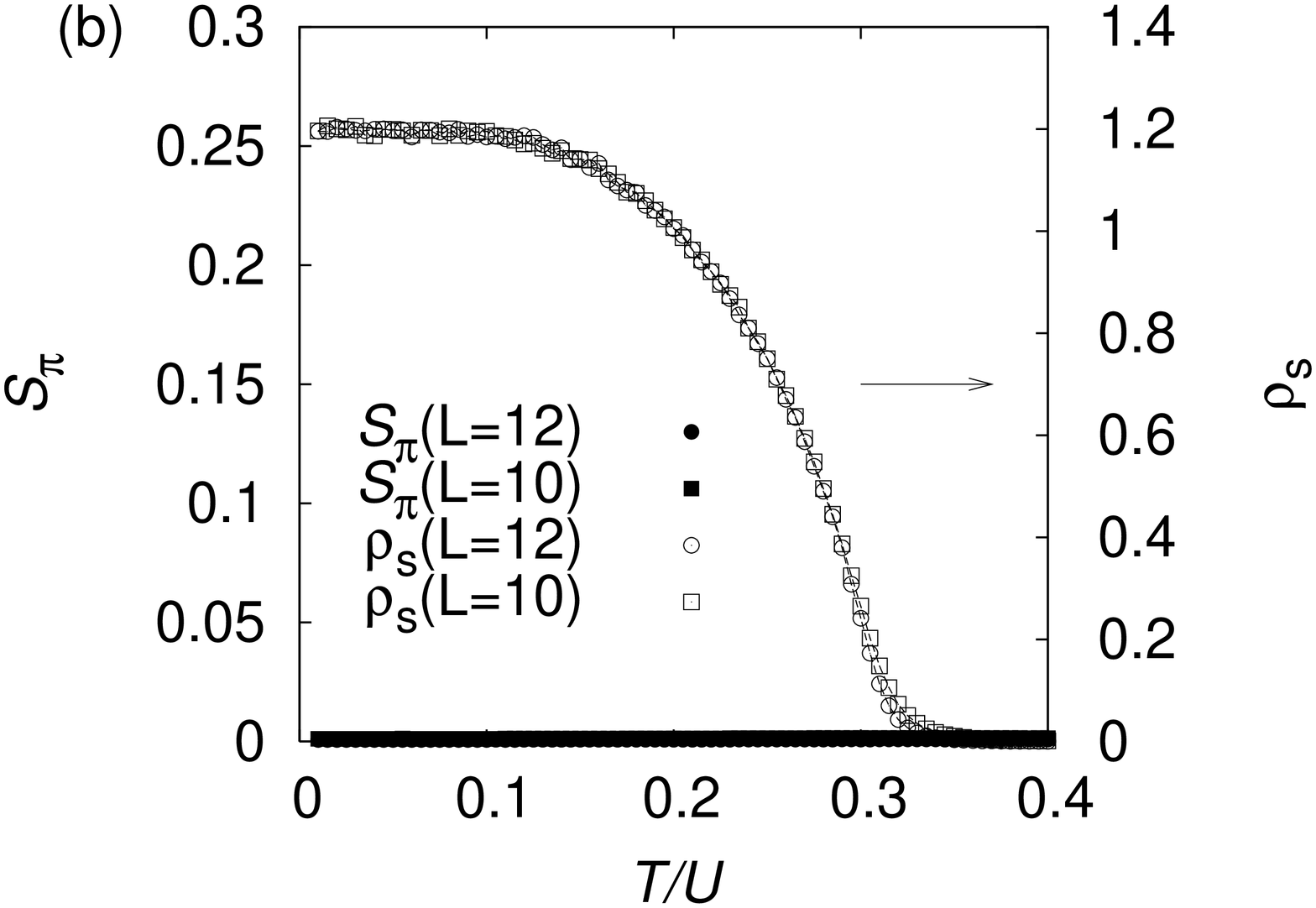}  \\
\includegraphics[width=5cm,angle=270]{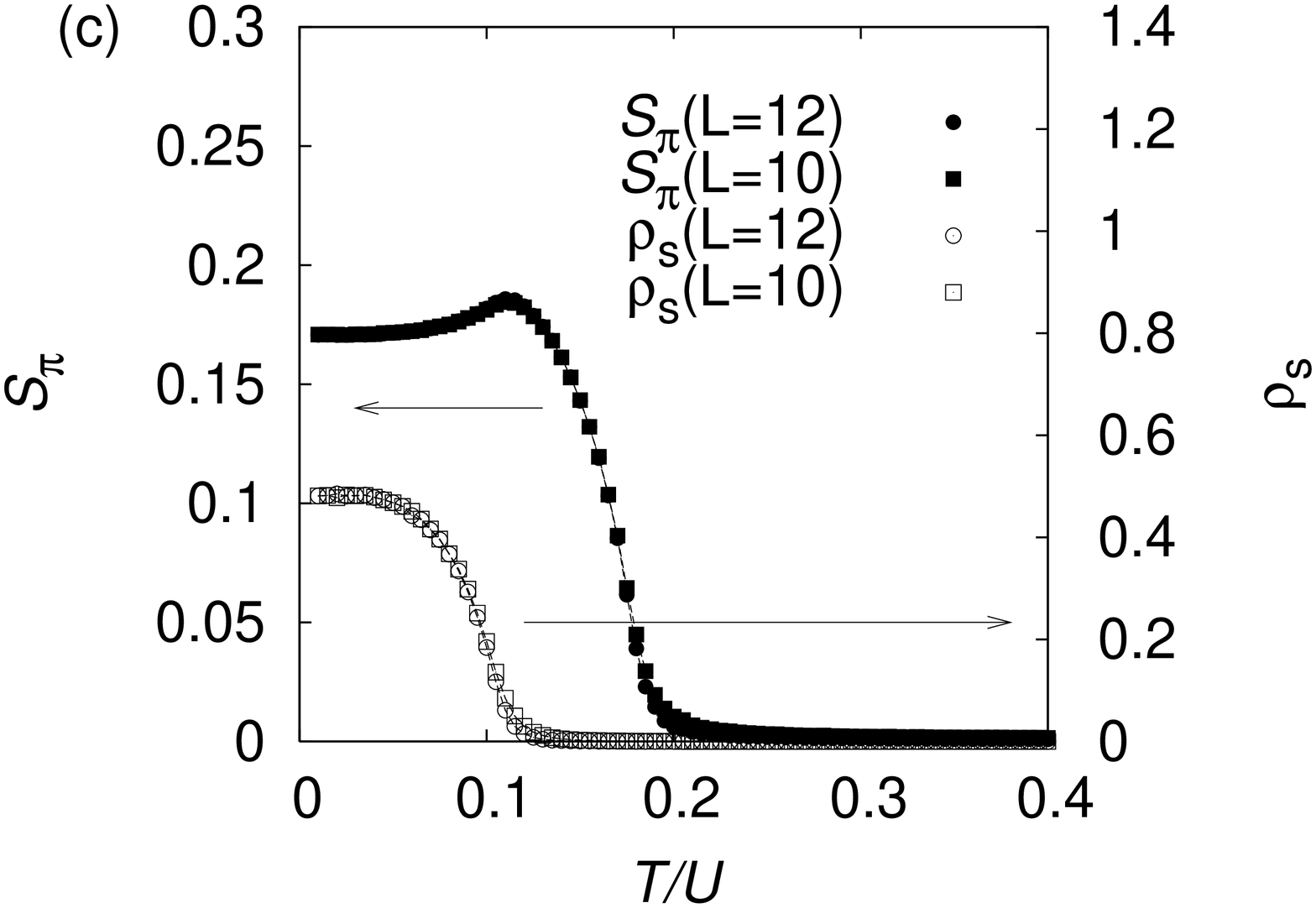} 
\caption{\label{fig-CBSFT} Temperature dependence of order parameters obtained by SSE ($V=U/z$).
 (a) $(t/U=0.02, \mu/U=0.7)$: Normal liquid - Normal solid transition. 
 (b) $(t/U=0.08, \mu/U=0.7)$: Normal liquid - Superfluid transition. 
 (c) $(t/U=0.045, \mu/U=0.7)$: Normal liquid - Normal solid transition and  Normal solid - Supersolid transition. 
}
\end{center}
\end{figure}
We plot the order parameters of solid, $S_{\pi}$, and that of superfluid, $\rho_s$, as a function of temperature for various values of $t$.
In Fig. \ref{fig-CBSFT} (a), we show the transition from normal liquid to normal solid for $t/U=0.02$ in which only $S_{\pi}$ appears.
In the same way, the transition from normal liquid to superfluid for $t/U=0.08$ is depicted in Fig. \ref{fig-CBSFT} (b).
For $t/U=0.045$, the system shows successive transitions and the supersolid state is realized at low temperatures.  
There, we find that the solid order appears at a higher temperature (Fig. \ref{fig-CBSFT} (c)).
Note that the solid order is suppressed when the superfluid order appears.
Thus, we expected that the solid fraction and the superfluid fraction compete with each other.
It should be noted that $\rho_s$ appears at higher temperature than the solid order for $t/U=0.55$ (not shown). 

{In Fig.~\ref{fig-SSEtTphase}, we depict a phase diagram in the coordinate of $(t/U,T/U)$ for the fixed values, $V/U=1/z$ and $\mu/U=0.7$.
The transition temperatures of the solid state $T_{\mathrm{S_{\pi}}}$ are plotted by solid circles, 
and those of the superfluid state $T_{\mathrm{\rho_s}}$ are plotted by open circles. 
To determine the transition temperatures for each value of $t$, we use the method of the Binder parameter of the systems with $L$=10 and 12.
In Fig.~\ref{fig-SSEtTphase}, there are four different phases: NL, NS, SF, and SS.
These phases meet at a tetra-critical point, ($t_c$, $T_c$). 
The competition of solid and superfluid orders is also found in the phase diagram (Fig. \ref{fig-SSEtTphase}).
Namely, above $t_c$, the transition temperature of solid is smaller than that of smooth extension of $T_{\rho_s}$ from $t<t_c$.
Therefore we conclude that $T_{\rho_s}$ is suppressed from that of the case in which the superfluid would not order.
Similarly, below $t_c$, the transition temperature of superfluid is smaller than that of the case in which the solid would not order. 

\begin{figure}
\includegraphics[width=6cm,angle=270]{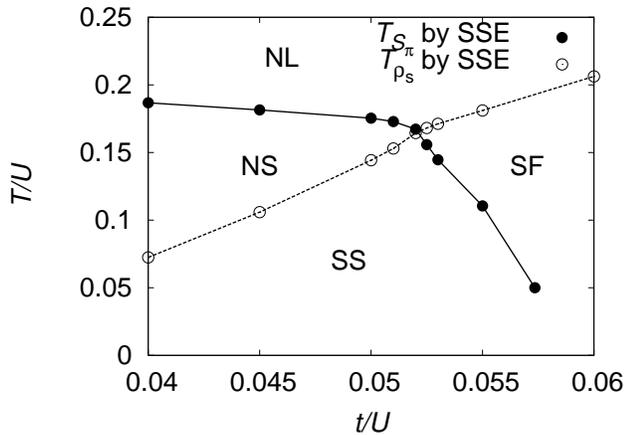}
\caption{\label{fig-SSEtTphase} The $t-T$ phase diagram for $V/U=1/z$ and $\mu/U=0.7$ obtained by SSE.
The transition temperatures of the solid state $T_{\mathrm{S_{\pi}}}$ are plotted by solid circles, 
and those of the superfluid state $T_{\mathrm{\rho_s}}$ are plotted by open circles. 
The lines connect the data points for the guide of eye.}
\end{figure}

\subsection{Mean-Field Analysis}
Here, we calculate the temperature dependence of order parameters by making use of  MF.
In Fig. \ref{fig-meanCBSFT}, we show the successive transitions of superfluid and solid for ($t/U=0.045, \mu /U=0.7$).
As was seen in the SSE simulation, here we find again the suppression of the solid order by the superfluid fraction. Namely, $S_{\pi}$ has a cusp
 at the superfluid transition point.
We also depict the phase diagram and compare that to that of SSE (Fig. \ref{fig-meantTphase}).
They show a qualitatively good agreement, e.g., 
there is the tetra-critical point $(t_c, T_c)$, and the critical temperatures $T_{\mathrm{S_{\pi}}}$ and $T_{\mathrm{\rho_s}}$ are suppressed by appearance of the other order,
as mentioned in the last session.
\begin{figure}
\includegraphics[width=5cm,angle=270]{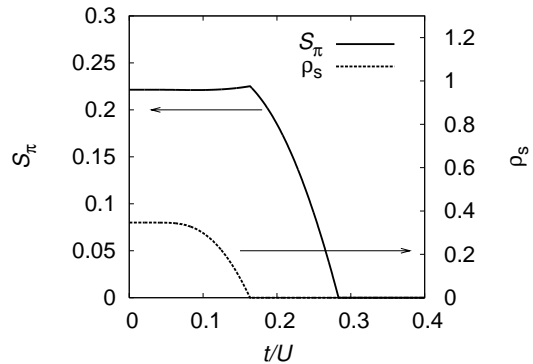}
\caption{\label{fig-meanCBSFT} Temperature dependence of order parameters for $V/U=1/z$, $t/U=0.045$, and $\mu/U=0.7$ obtained by MF.
The solid line denotes the solid order parameter and the dashed line denotes the superfluid order parameter.}
\end{figure} 
\begin{figure}
\includegraphics[width=5cm,angle=270]{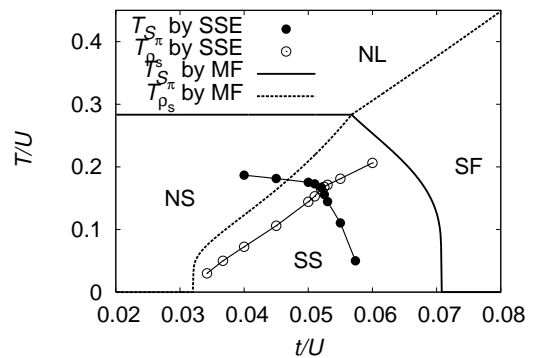}
\caption{\label{fig-meantTphase} The $t-T$ phase diagram for $V/U=1/z$ and $\mu/U=0.7$ obtained by MF. 
The transition temperature of the solid state $T_{\mathrm{S_{\pi}}}$ is denoted by the thick solid line, 
and that of the superfluid state $T_{\rho_S}$ is denoted by the thick dashed line.
We also depict the phase diagram (solid and open circles) obtained by SSE (Fig. \ref{fig-SSEtTphase}) for comparison.}
\end{figure}

Let us study the competition between the solid and superfluid orders by analyzing the Ginzburg Landau (GL) free energy.
Since the order parameters $m$ and $\phi$ take small value in the vicinity of the tetra-critical point, the GL free energy is expressed as 
\begin{equation}
F=am^2 +bm^4+c\phi^2+ d\phi^4+hm^2 \phi^2. 
\label{eq-GLfree}
\end{equation}
When $a$ becomes zero at $T_{\mathrm{S}_{\pi}}$ ( $a \simeq a_0(T-T_{\mathrm{S_{\pi}}})$) with a positive $b$, 
the second-order transition between the normal solid and normal liquid phase takes place, 
and similarly when $c$ becomes zero at $T_{\rho_s}$ ($c \simeq c_0 (T-T_{\mathrm{\rho_s}})$) with a positive $d$, 
the second-order transition between the superfluid and normal liquid phase takes place.  
The fifth term represents the competition between the solid and the superfluid order.
If $h$ equals zero, the transition of the solid phase
and that of superfluid take place independently at $T_{\mathrm{S_{\pi}}}$ and  $T_{\mathrm{\rho_s}}$, respectively.

If $h$ is positive, the presence of the superfluid order lowers the transition temperature of solid.
Below the $t_c$, the solid order emerges first as temperature decreases ($T_{\mathrm{S_{\pi}}}$).
Then, the superfluid order appears at the modified transition temperature $T_{\rho_s}^{\prime}$ which is smaller than the original one (Eq.(\ref{eq-TCBdecrease})).
\begin{eqnarray}
T_{\rho_s}^{\prime} < T_{\rho_s}-\frac{a_0 h/2b c_0} {1-a_0 h/2b c_0}(T_{\mathrm{S_{\pi}}}-T_{\rho_s}) < T_{\rho_s} \label{eq-TCBdecrease}
\end{eqnarray} 
In the same way, the solid order lowers the transition temperature of superfluid.
The changes of the transition temperature become large when $h$ becomes large, and this means the shrinkage of the supersolid region. 
Thus, $h$ represents the competition between the solid and the superfluid orders.  

The coefficients in the GL free energy can be calculated from the MF Hamiltonian, Eq. (\ref{eq-meanham}).
For example, the coefficient $a$ is expressed as
\begin{eqnarray}
a = (z-1)V-(z-1)^2 V^2 /{\cal Z_{\mathrm{MF}}} \nonumber \label{eq-coeffa} \\
\times Tr\left( \int_0^{\beta} d \lambda e^{\lambda {\cal H}_{\mathrm{MF}}} (n_{\capa} -n_{\capb}) e^{-\lambda {\cal H}_{\mathrm{MF}}} (n_{\capa} -n_{\capb}) \right).  
\end{eqnarray}
which is proportional to $(T-T_{\mathrm{S_{\pi}}})$ near the critical
point.

As a final part of this section, we discuss the effect of on-site repulsion on the coexistence of solid and superfluid orders.
As has been mentioned, the SS phase does not exist in the hardcore case. Since the hardcore is the limiting case of infinite $U$, we expect that the supersolidity is suppressed when the on site repulsion $U$ becomes large. 
We depict the $t-T$ phase diagram for various values of $U$ 
in Fig.~\ref{fig-SSUdep} obtained by MF. 
Here, we use $zV$ as the unit of energy instead of $U$ because now we want to study the effect of $U$. 
In Fig.~\ref{fig-SSUdep}, we find that the SS region becomes narrower as $U$ increases. 
Finally, the supersolid region disappears completely in the hardcore limit.
Thus, we conclude $U$, i.e. the hardness of the particle, suppresses the coexistence of the two orders.
\begin{figure}
  \begin{center}
    \begin{tabular}{cc}
      \resizebox{45mm}{!}{\includegraphics[angle=270]{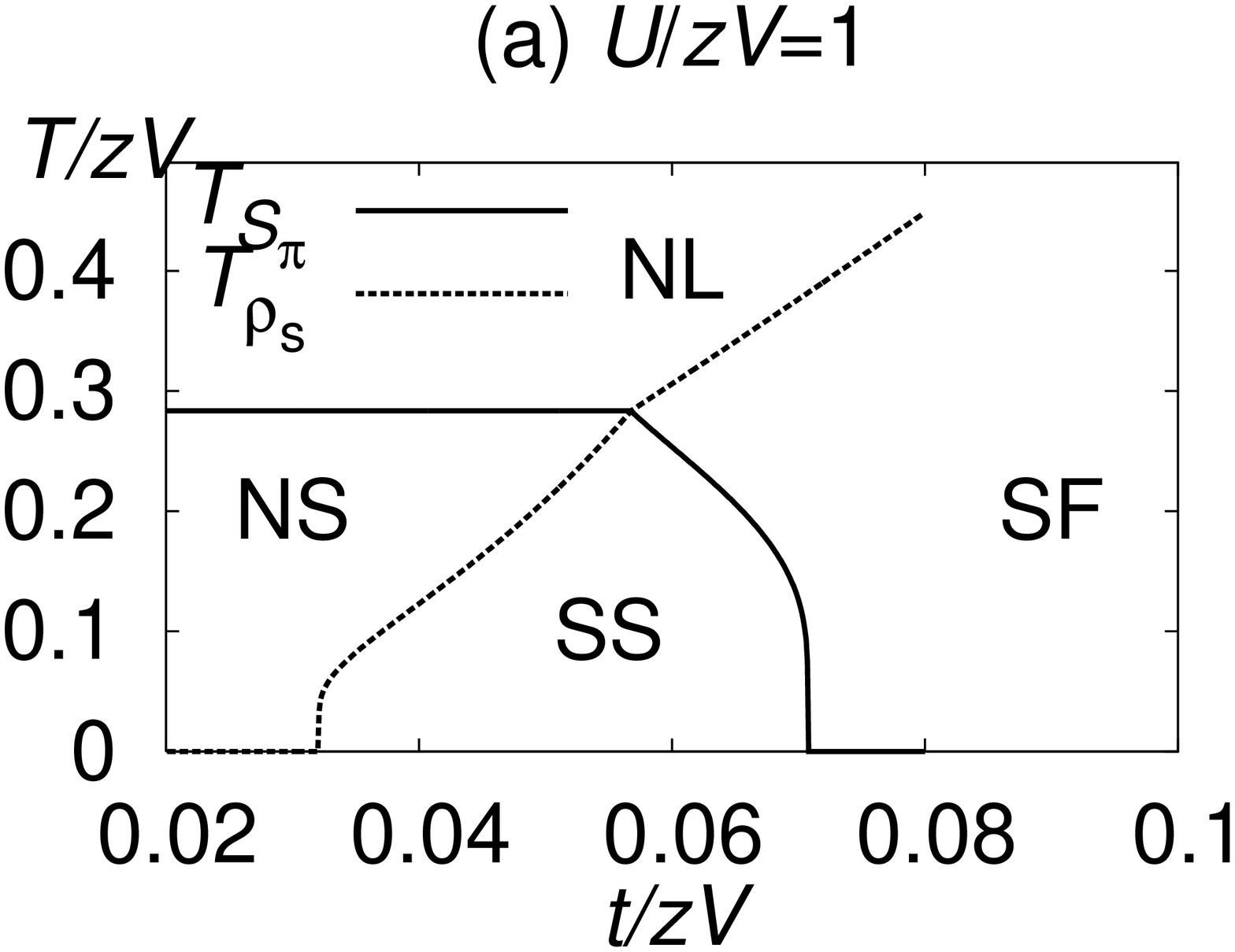}} &
      \resizebox{45mm}{!}{\includegraphics[angle=270]{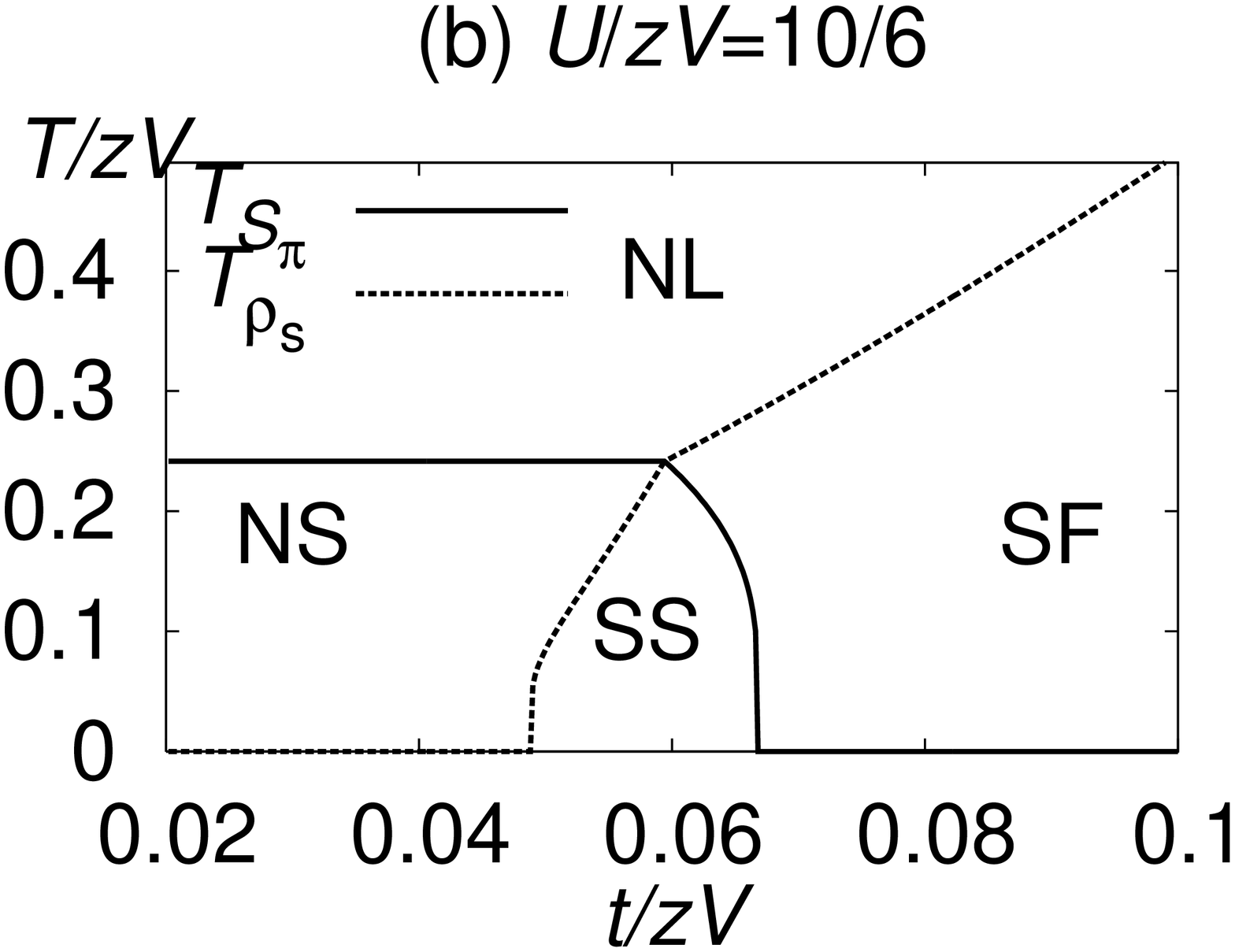}} \\      
      \resizebox{45mm}{!}{\includegraphics[angle=270]{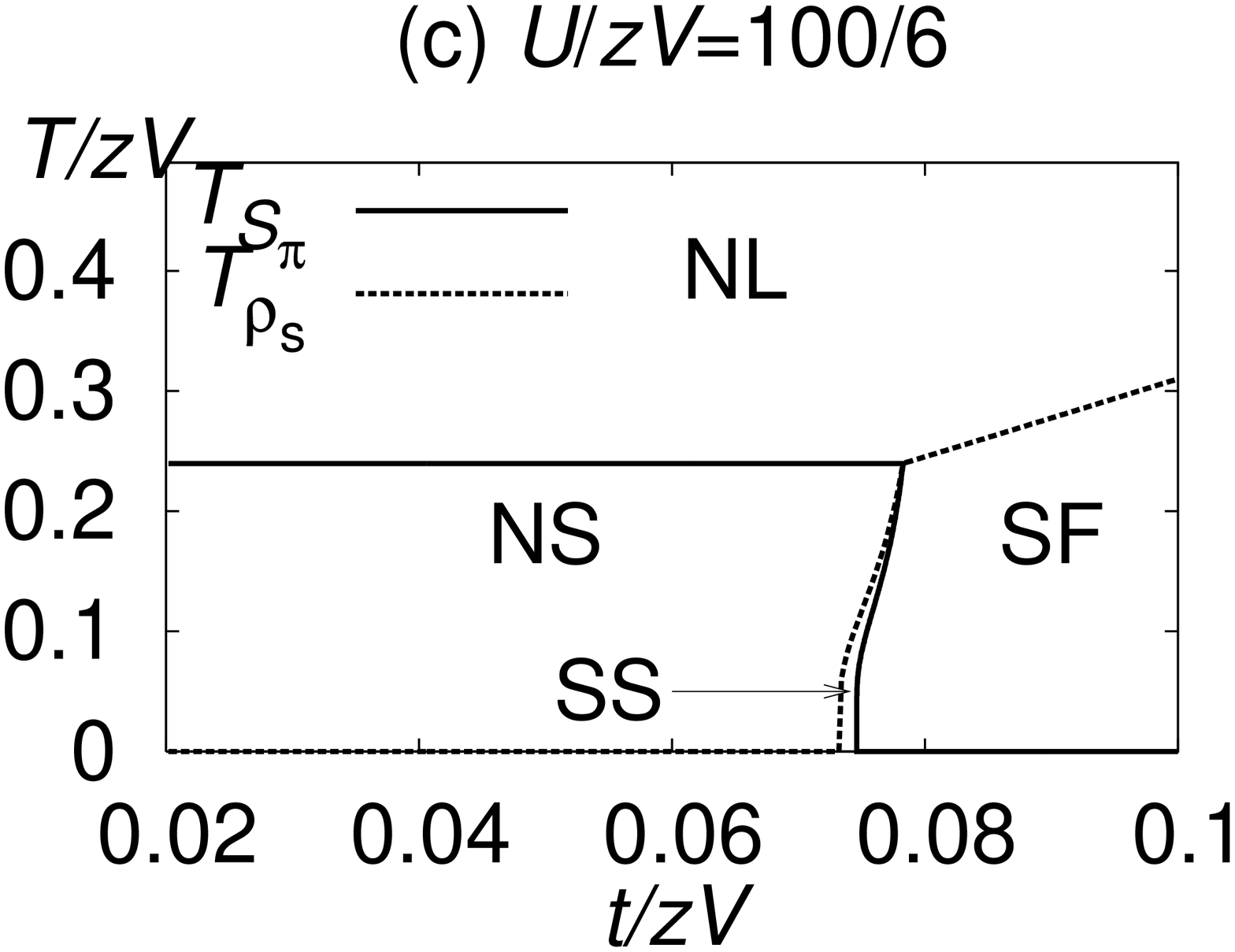}} &
      \resizebox{45mm}{!}{\includegraphics[angle=270]{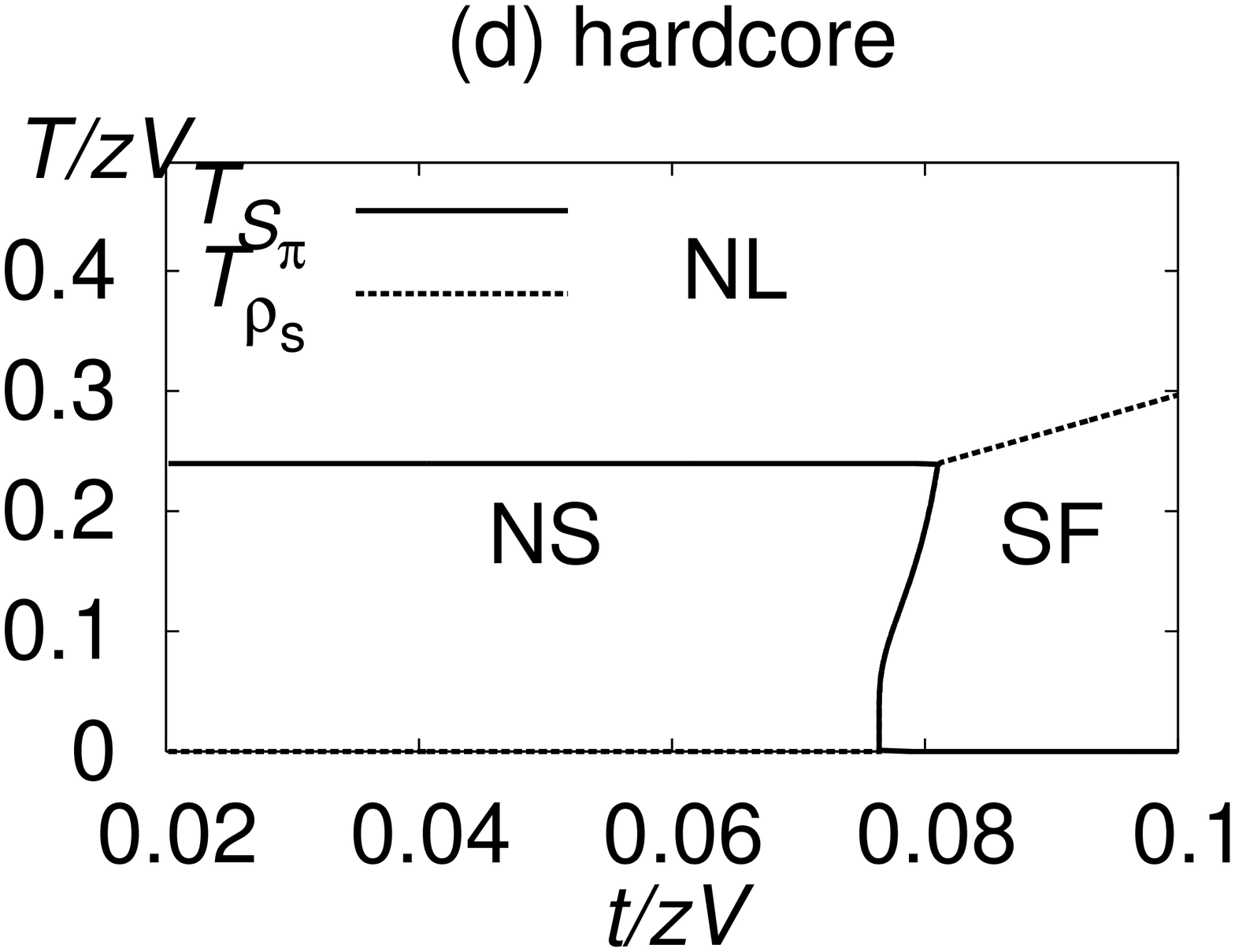}} \\
    \end{tabular}
    \caption{\label{fig-SSUdep}Finite temperature phase diagrams obtained by MF for various values of $U/V$. 
(a) $U/V=6$, (b) $U/V=10$, (c) $U/V=100$, (d) hardcore ($U/V=\infty$).}
  \end{center}
\end{figure}


\section{discussion and summary} 
By using SSE simulation and MF analysis, we study successive phase transitions of the solid order and superfluidity at finite temperatures
 in the case the system has the supersolid state in the ground state.
  The strong hopping term (large $t$) favors the superfluid and the superfluid phase appears at
 higher temperature in the large $t$ region. On the other hand, in the small $t$ region, the normal solid phase appears first.  In both cases
we find that the orders of superfluid and solid appear at different temperatures, i.e., successive phase transitions.
We find a phase diagram with a tetra-critical point and studied the competition between the solid and superfluid orders. 
By analyzing the dependence of the phase diagram on the on-site repulsion $U$ in particular near the tetra-critical point, 
we find that larger $U$ enhances the competition between two orders and causes the shrinkage of the supersolid region. 
It would help us to understand how the softness of particles contributes to the realization of the supersolid.

Possibility of realization of the supersolid state on the optical lattice has been discussed recently. 
The realization of the Bose-Hubbard model in the optical lattice has been discussed \cite{jaksch1998}.
For realization of the supersolid state, the nearest neighbor repulsive interaction $V$ plays an important role.  
Mazzarella, et al. discussed how to introduce the nearest neighbor interaction \cite{mazzarella2006}.
We expect that the parameters of the system can be widely controlled in the optical lattice and there
the properties of the phase diagram obtained in this paper will be observed. 


\section*{Acknowledgments}
\label{sec:ACK}

The authors would like to thank Professor Hiroshi Fukuyama 
for his valuable discussions.
We also acknowledge useful discussions with Keigo Hijii.
The numerical simulations were performed using ALPS applications and libraries \cite{alps2}.
This work was partially supported by a Grant-in-Aid for Scientific 
Research on Priority Areas
``Physics of new quantum phases in superclean materials" 
(Grant No.\ 17071011), 
and also by the Next Generation Super Computer Project, 
Nanoscience Program of MEXT.
Numerical calculations were done on the supercomputer of ISSP.
\bibliographystyle{junsrt}

\begin{thebibliography}{10}

\bibitem{penrose1956}
O.~Penrose and L.~Onsager.
\newblock {\em Phys. Rev.}, Vol. 104, p. 576, 1956.

\bibitem{kim2004}
E.~Kim and M.~H.~W. Chan.
\newblock {\em SCIENCE}, Vol. 305, p. 1941, 2004.

\bibitem{sasaki2006}
S.~Sasaki, R.~Ishiguro, F.~Caupin, H.~J. Maris, and S.~Balibar.
\newblock {\em SCIENCE}, Vol. 313, p. 1098, 2006.

\bibitem{andreev1969}
A.~F. Andreev and I.~M. Lifshitz.
\newblock {\em Sov. Phys. JETP}, Vol.~29, p. 1107, 1969.

\bibitem{matsuda1970}
H.~Matsuda and T.~Tsuneto.
\newblock {\em Suppl. Prog. Theor. Phys.}, Vol.~46, p. 411, 1970.

\bibitem{boninsegni2003}
M.~Boninsegni.
\newblock {\em Jour. Low Temp. Phys}, Vol. 132, p.~39, 2003.

\bibitem{batrouni2000}
G.~G. Batrouni and R.~T. Scalettar.
\newblock {\em Phys. Rev. Lett}, Vol.~84, p. 1599, 2000.

\bibitem{suzuki2007}
T.~Suzuki and N.~Kawashima.
\newblock {\em Phys. Rev. B}, Vol.~75, p. 180502, 2007.

\bibitem{sengupta2005}
P.~Sengupta, L.~P. Pryadko, F.~Alet, M.~Troyer, and G.~Schmid.
\newblock {\em Phys. Rev. Lett.}, Vol.~94, p. 207202, 2005.

\bibitem{sandvik1991}
A.~W. Sandvik and J.~Kurkij{\"a}rvi.
\newblock {\em Phys. Rev. B}, Vol.~43, p. 5950, 1991.

\bibitem{sandvik1999}
A.~W. Sandvik.
\newblock {\em Phys. Rev. B}, Vol.~59, p. 14157, 1999.

\bibitem{alet2005}
F.~Alet, S.~Wessel, and M.~Troyer.
\newblock {\em Phys. Rev. E}, Vol.~71, p. 036706, 2005.

\bibitem{alps2}
http://dx.doi.org/10.1016/j.jmmm.2006.10.304.

\bibitem{pollock1987}
E.~L. Pollock and D.~M. Ceperley.
\newblock {\em Phys. Rev. B}, Vol.~36, p. 8343, 1987.

\bibitem{prokofev2000}
N.~V. Prokof'ev and B.~V. Svistunov.
\newblock {\em Phys. Rev. B}, Vol.~61, p. 11282, 2000.

\bibitem{xiancong2006}
X.~Lu and Y.~Yu.
\newblock {\em Phys. Rev. A}, Vol.~74, p. 063615, 2006.

\bibitem{otterlo1995}
A.~van Otterlo, K.~H. Wagenblast, R.~Baltin, C.~Bruder, R.~Fazio, and
  G.~Sch{\"o}n.
\newblock {\em Phys. Rev. B}, Vol.~52, p. 16176, 1995.

\bibitem{YamamotoMiyashita1}
K.~Yamamoto and S.~Miyashita.

\bibitem{jaksch1998}
D.~Jaksch, C.~Bruder, J.~I. Cirac, C.~W. Gardiner, and P.~Zoller.
\newblock {\em Phys. Rev. Lett.}, Vol.~81, p. 3108, 1998.

\bibitem{mazzarella2006}
G.~Mazzarella, S.~M. Giampaolo, and F.~Illuminati.
\newblock {\em Phys. Rev. A}, Vol.~73, p. 013625, 2006.

\end{thebibliography}

\end{document}